\documentclass[sigconf,nonacm]{acmart}
\usepackage{amsmath,amsfonts}
\usepackage{algorithmic}
\usepackage{graphicx}
\usepackage{textcomp}
\usepackage{xcolor}
\usepackage{booktabs}
\usepackage{tabularx}

\def\BibTeX{{\rm B\kern-.05em{\sc i\kern-.025em b}\kern-.08em
    T\kern-.1667em\lower.7ex\hbox{E}\kern-.125emX}}
\usepackage{hyperref}

\newif\ifshownotes
\shownotestrue

\usepackage[most]{tcolorbox}
\newtcolorbox{Finding}{
    colback=yellow!20,
    colframe=yellow!20,
    boxrule=0pt,
    sharp corners,
    boxsep=0pt,
    left=8pt,
    right=8pt,
    top=6pt,
    bottom=6pt,
    width=\linewidth,
    before skip=8pt,
    after skip=8pt}

\begin{document}

\title{Requirements Nobody Wrote Down: Emergence in a Public-Sector ML Pipeline}


\author{Ronivaldo Ferreira}
\affiliation{%
  \institution{Federal University of Par\'a}
  \city{Bel\'em}
  \state{PA}
  \country{Brazil}}
\email{ronivaldo.junior@icen.ufpa.br}

\author{Vinicius Costa Girard}
\affiliation{%
  \institution{Federal University of Par\'a}
  \city{Bel\'em}
  \state{PA}
  \country{Brazil}}
\email{vinicius.girard@icen.ufpa.br}

\author{Cleidson de Souza}
\affiliation{%
  \institution{Federal University of Par\'a}
  \city{Bel\'em}
  \state{PA}
  \country{Brazil}}
\email{cdesouza@ufpa.br}

\author{Igor Steinmacher}
\affiliation{%
  \institution{Northern Arizona University}
  \city{Flagstaff}
  \state{AZ}
  \country{USA}}
\email{igor.steinmacher@nau.edu}

\author{Carla Aguiar}
\affiliation{%
  \institution{University of Bras\'{\i}lia (UnB)}
  \city{Bras\'{\i}lia}
  \state{DF}
  \country{Brazil}}
\email{caguiar@unb.br}

\author{Filipe Roseiro Cogo}
\affiliation{%
  \institution{Queen's University}
  \city{Kingston}
  \state{ON}
  \country{Canada}}
\email{filipe.cogo@queensu.ca}

\author{Gustavo Pinto}
\affiliation{%
  \institution{Federal University of Par\'a}
  \city{Bel\'em}
  \state{PA}
  \country{Brazil}}
\email{gpinto@ufpa.br}

\renewcommand{\shortauthors}{Ferreira et al.}

\begin{abstract}
Governments increasingly use machine learning (ML) to handle data volumes beyond their analytical capacity, yet little is known about how requirements arise and evolve as an ML experiment moves toward operational use. Requirements Engineering (RE) for ML often assumes that requirements can be elicited upfront; in the public sector, however, demands arrive vague, stakeholders speak different professional languages, and what counts as a ``good'' result is unknown in advance. We report a 15-month case study of a thematic classification pipeline for Brazil's national digital participation platform, which holds over 60,000 citizen proposals, triangulating 43 project artifacts with interviews of six project participants. Institutional demands arrived as broad goals, endorsed but not sponsored, and only 40\% of the 95 reconstructed requirements and criteria were ever formalized. The pipeline's key requirements emerged largely from experiments, model outputs, and integration work, shaped by technical decisions and institutional constraints; the central one, the classification categories, was settled by a government taxonomy chosen for its institutional legitimacy. Acceptance relied on technical metrics, domain judgment, and political assessment, without shared thresholds or validation of usefulness in practice. In public-sector ML, technical and institutional artifacts thus act as elicitation instruments, and experimentation and integration belong to the requirements process.
\end{abstract}


\maketitle

\section{Introduction}
Machine learning (ML) is making its way into the public sector. Governments use it to triage citizen requests, detect fraud, and analyze text volumes that would take months to read manually~\cite{wirtz2019}. Digital participation platforms are a prominent case: by opening consultations to any citizen, they collect thousands of free-text contributions that agencies must read, organize, and answer. Brazil's national platform has received over 60,000 proposals across 117 participatory processes and two assemblies since 2023, on topics from national planning to climate and cultural policy. At this scale, manual analysis exceeds the agencies' capacity, making ML a precondition for taking citizen input seriously.

Building such solutions is not only a modeling problem but a requirements problem. A request such as ``classify the proposals by theme'' leaves open what counts as a theme, whose vocabulary the themes should follow, how good the classification must be before managers rely on it, and who judges it. In conventional software, Requirements Engineering (RE) answers these questions by eliciting and specifying stakeholder needs before or alongside implementation. ML-enabled systems challenge this practice: their behavior is learned from data, their quality is expressed in metrics whose thresholds are rarely known in advance, and their requirements depend on data that may not yet exist or be labeled~\cite{vogelsang2019,horkoff2019}. These differences reshape software engineering practice~\cite{amershi2019} and create friction between data scientists and their teams~\cite{nahar2022}.

The RE for ML community has characterized how data scientists perceive requirements~\cite{vogelsang2019}, mapped the challenges of non-functional requirements~\cite{horkoff2019}, studied RE in safety-critical perception systems~\cite{habibullah2023}, and identified gaps in human-centered AI~\cite{ahmad2023}; secondary studies have mapped the field~\cite{villamizar2021}. This literature has two blind spots. First, most studies capture RE as practitioners \emph{report} it at one point in time~\cite{vogelsang2019,alves2023}, rarely following a system long enough to see its requirements \emph{emerge, shift, and become operational}. Second, the public sector is largely absent. There, demands arrive through administrative instruments, stakeholders come from different professions, and success is defined by institutional accountability.


Consider the thematic classifier we studied. The team first clustered proposals with topic modeling, and no requirement named a vocabulary. When the team asked for one, the Controlled Vocabulary of Electronic Government (VCGE), a taxonomy used across the federal administration, was suggested, and the team began seeding the model with its terms. What started as a point of comparison became the criterion for judging topics and helped motivate supervised classification, so that categories matched the nomenclature managers use to report on public policy. No requirements document recorded this shift; it surfaced in scripts, presentations, and meeting notes. Such episodes suggest that, in public-sector ML, requirements may be \emph{discovered} through experimentation and institutional artifacts as much as elicited from stakeholders.

We report a qualitative case study of this pipeline over 15 months (January 2025 to March 2026). Rather than assume that requirements existed upfront, we ask how they came into being and what role they played:

\smallskip
\noindent\textbf{RQ1: How did institutional demands become requirements for the ML pipeline?} We examine where requirements originated, who formulated them, and whether they were elicited upfront or surfaced over time.

\smallskip
\noindent\textbf{RQ2: How did requirements and technical decisions shape each other over time?} We ask whether requirements drove the main technical decisions, emerged from them, or both.

\smallskip
\noindent\textbf{RQ3: How was the acceptability of the pipeline's results established?} We examine which criteria were used, who applied them, and how technical metrics, domain judgment, and institutional concerns combined.

\smallskip
We first analyzed all 43 project artifacts, normalizing 95 requirements and criteria and reconstructing a trajectory of 17 episodes. We then interviewed six participants, using the episodes to prompt recall and challenge our reconstruction, coded the transcripts deductively and inductively, and consolidated the episodes. The resulting seven patterns, triangulated with the episodes, answer our RQs.\footnote{Replication package: \href{https://doi.org/10.6084/m9.figshare.34867212}{\nolinkurl{doi.org/10.6084/m9.figshare.34867212}}.}

Our main findings are:

\begin{itemize}
\item Requirements existed but were rarely written down: only 40\% of the 95 requirements and criteria we reconstructed were ever formalized.
\item Institutional demands arrived as broad goals. AI was the team's offer, endorsed but never sponsored, and the team itself wrote down its requirements.
\item Most requirements emerged from experiments, model outputs, and integration after modeling had started. The central one, the classification categories, had no owner and was settled by a government taxonomy chosen for legitimacy and distant from citizens' language.
\item Acceptance relied on technical metrics, domain judgment, and political assessment, with no shared threshold and no evaluation in use.
\end{itemize}



\section{Study Context}
\label{sec:context}

Brazil's national digital participation platform runs on Decidim, an open-source platform for participatory democracy~\cite{barandiaran2024}. Federal agencies use it for participatory processes on public policies such as the Multi-Year Plan (PPA), the Climate Plan, and the National Culture Plan. In each process, the agency opens a consultation, citizens and organizations submit and vote on proposals, managers analyze them, and the agency responds formally to society. By our count of the database, over 60,000 proposal records had accumulated across 117 processes and two assemblies by March 2026, before cleaning. For comparison, a single planning process received 8,254 proposals~\cite{jardim2024ppa}, and a semantic clustering study analyzed 10,022 proposals from three processes~\cite{ferreira2025semantic}.

\subsection{Actors and ML Initiatives}

\begin{figure}[t]
\centering
\includegraphics[width=\columnwidth]{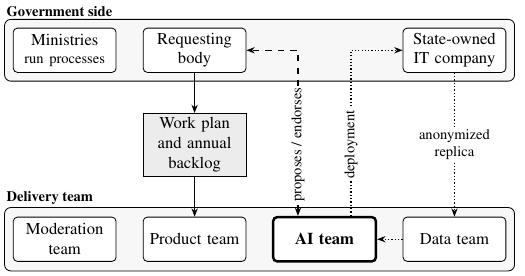}
\caption{Actors and demand paths. Solid: formal path from the requesting body through the work plan to the product team. Dashed: AI fronts proposed by the delivery team and endorsed outside the work plan. Dotted: the AI team's data and deployment dependencies.}
\label{fig:actors}
\end{figure}

On the government side, a national body for social participation requests the work and owns the platform, ministries run the processes and receive the contributions, and a state-owned IT company hosts the infrastructure and data. On the delivery side, a team funded through a public-sector cooperation agreement (a legal instrument by which a government body funds another public entity, such as a university, to carry out a work plan) develops and maintains the platform. It has a \emph{product team}, the main interface with the requesting body; a \emph{data team}, which extracts data from an anonymized replica of the production database; an \emph{AI team} of data scientists, which builds models on these data; and a \emph{moderation team} of political scientists and lawyers, which moderates content and analyzes proposals thematically. AI work was not in the work plan: the delivery team proposed the AI fronts, which were endorsed outside the agreed deliverables (Section~\ref{sec:pa}). Fig.~\ref{fig:actors} shows these actors and how demands and dependencies reached the AI team, the team we studied.


The AI team was created to explore how ML could support the participatory cycle, starting from proposals written in citizens' own words, often far from the administration's vocabulary. Its initial goals were to flag content that violates the terms of use, classify proposals by theme so managers could systematize them and produce reports and formal responses, and recommend proposals and processes to citizens. We focus on the thematic classifier, which served the last two goals and thus two potential consumers: managers and citizens. Over the study period, the team moved from topic modeling with BERTopic, to seeding it with VCGE terms, to supervised classification on an adaptation of this vocabulary, and finally to a service with inference, training, and model-publishing routines whose deployment depended on the state-owned company's infrastructure. We draw on other fronts, such as a retrieval-augmented generation (RAG) system for reports and a Climate Plan prototype built on commercial APIs, when they help explain the classifier.

\subsection{Researcher Positionality}
\label{sec:posicao}

This is an insider study. The first author has been an ML engineer in the AI team since January 2025, working on data preparation, model training, and validation. This position gave us access to participants and artifacts; some presentations and meeting notes came from the first author's files and were included without selection. It also means that the interviewer was a team member and that the first author analyzed decisions they took part in (Section~\ref{sec:threats}). To mitigate this, we kept reflective memos, let participants confirm, correct, or contest the reconstructed episodes, consolidated them under a protocol fixed in advance (Section~\ref{sec:consolidacao}), and had the second author review all coding and the candidate patterns (Section~\ref{sec:codificacao}).

\section{Methodology}
\label{sec:method}

\begin{figure*}[t]
    \centering
    \includegraphics[width=\textwidth]{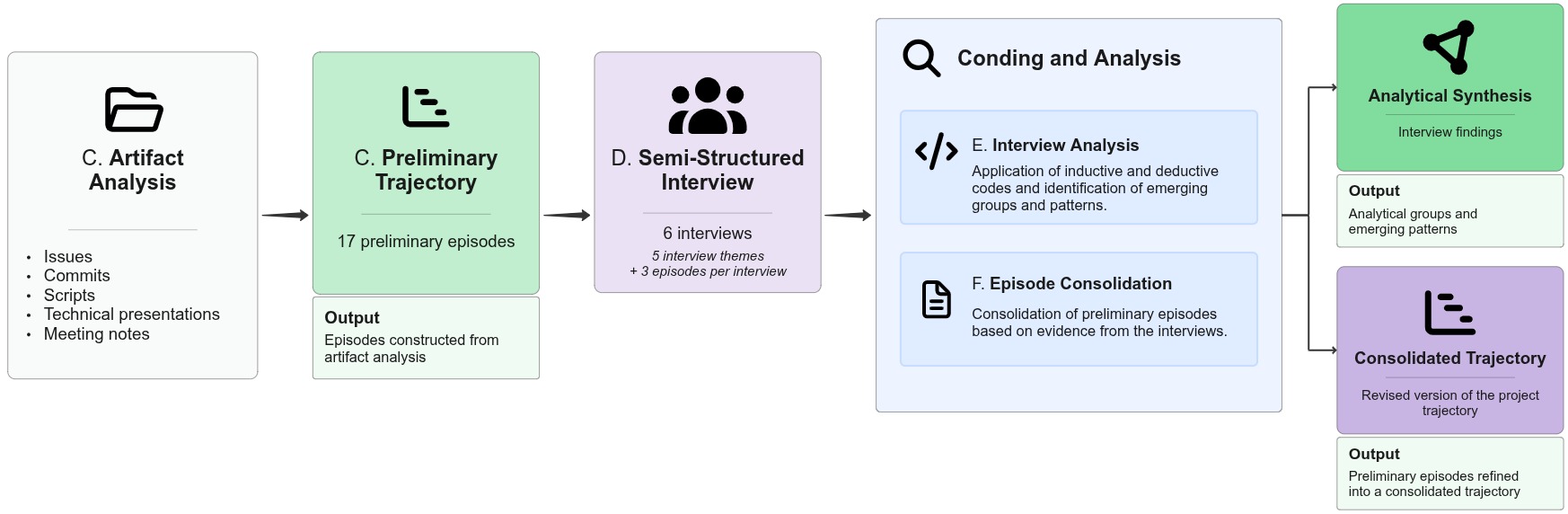}
    \caption{Research method: (1) artifact analysis and construction of 17 preliminary episodes; (2) six semi-structured interviews covering five themes, each supported by preliminary episodes; (3) deductive and inductive coding; and (4) consolidation of the episodes and synthesis of patterns.}
    \label{fig:method}
\end{figure*}

We conducted an exploratory single-case study~\cite{runeson2009,yin2018}. The \emph{case} is the thematic classification pipeline (Section~\ref{sec:context}); the \emph{unit of analysis} is the trajectory of its requirements and evaluation criteria from January 2025 to March 2026. The case is \emph{revelatory}~\cite{yin2018}: the first author's insider position (Section~\ref{sec:posicao}) gave us access to all project artifacts and to participants on both the delivery and government sides, which is rare for public-sector ML systems. The design is longitudinal in its unit of analysis and retrospective in data collection, with two stages (Fig.~\ref{fig:method}): we reconstructed a preliminary trajectory from the artifacts, then used it to support interviews and consolidated it with their evidence. Artifacts show what was recorded and when, since they were produced to do the work and dated before outcomes were known; interviews show the motivations, negotiations, and tacit expectations that the artifacts left out~\cite{runeson2009}.


\subsection{Operational Definition of Requirement}
\label{sec:req-def}
Lacking a specification, we needed a definition broad enough for requirements never written as such. A \emph{requirement} is any condition the solution had to satisfy to be accepted, integrated, or used, documented or not. \emph{Formalized} requirements were recorded as such in some project instrument; \emph{tacit} ones guided decisions without being recorded. Across both, we mark \emph{environmental constraints}: conditions imposed by hardware, contracts, or data governance that the team treated as requirements. Following RE for ML terminology~\cite{vogelsang2019,habibullah2023}, we separate functional (FR) and non-functional requirements (NFR) from three kinds of criteria: acceptance criteria (AC), stating when an output or deliverable was accepted; evaluation criteria (EC), stating how the solution was assessed; and success criteria (SC), stating what made the initiative worthwhile.

\subsection{Artifact Collection}
\label{sec:collection}
In June and July 2026, we collected every artifact produced for the pipeline during the study period, without filtering: 10 issues and 15 commits from the project's GitLab repository, 6 Google Colab notebooks, and 5 presentations and 7 meeting notes from the authors' files, 43 in total. We cataloged each with an identifier, type, date, description, and phase.

\subsection{Artifact Analysis and Preliminary Trajectory}
\label{sec:artifact-analysis}
We analyzed each artifact with guiding questions for its type: issues and scripts for demands, requirements, and criteria; commits, presentations, and meeting notes for the context, motivations, and results of technical decisions. We cross-referenced artifacts so that one could complete or corroborate another (e.g., an issue announces a task, a script implements it, and a presentation reports its results). The first author conducted this analysis alone.

\subsubsection{Normalizing requirements and criteria}
The same need often appeared in different words across artifacts and phases. We merged equivalent formulations into canonical items, each with an identifier, a definition, and links to its original formulations and artifacts; for example, ``saving'', ``publishing'', and ``loading'' the model became the FR \emph{Manage the model lifecycle and availability}. We created a new item only when a formulation fit no existing one, yielding 21 FRs, 12 NFRs, 23 ACs, 22 ECs, and 17 SCs.


\subsubsection{Preliminary episodes}
\label{sec:episodes}
A \emph{preliminary episode} (EP) is a significant delivery, turning point, or change of approach or tooling where normalized requirements or criteria first appeared, reappeared, were refined, or were operationalized. For each EP, we recorded the demand, requirements, criteria, technical decisions, results, and supporting artifacts, so the same requirement can be followed across episodes. This produced 17 EPs in six phases: (1)~exploration and feasibility; (2)~unsupervised modeling and categorization; (3)~validation, alignment, and scale; (4)~transition to supervised classification; (5)~operationalization and deployment; and (6)~pipeline governance. For example, EP07, \emph{VCGE becomes the institutional reference}, links an issue requesting high-level categories, a script seeding BERTopic with VCGE terms, and two presentations of results to five requirements and seven criteria.


\subsection{Semi-Structured Interviews}
\label{sec:interviews}

\paragraph{Participants}
Through purposive sampling~\cite{runeson2009}, we interviewed the six stakeholders most involved with the pipeline, covering the origin of demands, their translation into requirements, technical decisions, and evaluation (Table~\ref{tab:participants}): two political scientists, one of whom directed the requesting body; two software engineers; a data engineer; and an electrical engineer in ML leadership. For a single, bounded pipeline, we favored depth of involvement over the number of participants.


\begin{table}[t]
\centering
\caption{Interview participants from the delivery and government sides. Min.: interview length in minutes.}
\label{tab:participants}
\footnotesize
\setlength{\tabcolsep}{5pt}
\begin{tabular}{@{}llllr@{}}
\hline
\textbf{ID} & \textbf{Background} & \textbf{Role in the project} & \textbf{Side} & \textbf{Min.} \\
\hline
E01 & Political sci. & Product and domain & Delivery & 48 \\
E02 & Electrical eng. & ML leadership & Delivery & 73 \\
E03 & Political sci. & Director, req.\ body & Government & 66 \\
E04 & Software eng. & Product eng., AI team & Delivery & 92 \\
E05 & Software eng. & Delivery coordination & Delivery & 69 \\
E06 & Data eng. & Data engineering & Delivery & 64 \\
\hline
\end{tabular}
\end{table}

\paragraph{Protocol}
The first author conducted all interviews remotely, in Brazilian Portuguese, between August 21 and September 18, 2026 (mean: 68 minutes). Three pilot interviews, excluded from the analysis, refined the guide, which covered five themes: (1)~origin of the demand and the participant's role; (2)~from demands to requirements; (3)~from requirements to technical decisions; (4)~acceptance, evaluation, and success; and (5)~evolution, negotiation, and priorities. After the open questions, we showed three to five EPs chosen by the participant's role and asked them to confirm, complete, or correct each one (\emph{episode-supported elicitation}). We transcribed the interviews with WhisperX (Whisper large-v3), and the first author wrote a reflective memo after each.



\subsection{Interview Analysis}
\label{sec:codificacao}
We coded the interviews in two cycles~\cite{saldana2009coding}: deductive and inductive coding, then pattern coding that combined both. Interviews were always coded in the same order, starting with those richest in project context.


\subsubsection{Deductive codebook}
We derived 23 codes in six dimensions (Table~\ref{tab:deductive-codebook-short}) from four RE for ML works chosen for their complementary coverage: data scientists' practices~\cite{vogelsang2019}, safety-critical systems~\cite{habibullah2023}, human-centered AI~\cite{ahmad2023}, and NFRs~\cite{horkoff2019}. Each code has a definition, application criteria, and supporting quotes from its source.

\begin{table}
\centering
\caption{Deductive codebook: 23 codes in six dimensions,
  with the works each dimension draws on.}
\label{tab:deductive-codebook-short}
\footnotesize
\setlength{\tabcolsep}{4pt}
\newcommand{\dimrow}[2]{%
  \addlinespace[3pt]\textbf{#1} & #2 \\}
\newcommand{\coderow}[1]{%
  \multicolumn{2}{@{\hspace{1em}}%
  p{\dimexpr\columnwidth-1em\relax}@{}}%
  {\raggedright #1} \\}
\begin{tabular*}{\columnwidth}%
  {@{\extracolsep{\fill}}l r@{}}
\toprule
\textbf{Dimension and codes} & \textbf{Sources} \\
\midrule
\dimrow{D1. Requirement types}
  {\cite{vogelsang2019,habibullah2023,%
  ahmad2023,horkoff2019}}
\coderow{1.1~Quantitative targets
  \textperiodcentered{} 1.2~Explainability
  \textperiodcentered{} 1.3~Discrimination/bias
  \textperiodcentered{} 1.4~Legal/regulatory
  \textperiodcentered{} 1.5~Data/annotations}
\dimrow{D2. RE process}
  {\cite{vogelsang2019,habibullah2023}}
\coderow{2.1~Elicitation
  \textperiodcentered{} 2.2~Analysis
  \textperiodcentered{} 2.3~Specification
  \textperiodcentered{} 2.4~Continuous V\&V}
\dimrow{D3. Data scientists' challenges}
  {\cite{vogelsang2019,habibullah2023,ahmad2023}}
\coderow{3.1~Communication gap
  \textperiodcentered{} 3.2~Paradigm shift
  \textperiodcentered{} 3.3~Training vs.\ operational performance}
\dimrow{D4. RE topics for ML}
  {\cite{habibullah2023}}
\coderow{4.1~Scenarios
  \textperiodcentered{} 4.2~Requirements breakdown
  \textperiodcentered{} 4.3~Traceability}
\dimrow{D5. Human-centered AI}
  {\cite{ahmad2023}}
\coderow{5.1~User needs
  \textperiodcentered{} 5.2~Model needs
  \textperiodcentered{} 5.3~Feedback/control
  \textperiodcentered{} 5.4~Graceful failure
  \textperiodcentered{} 5.5~Tool gap
  \textperiodcentered{} 5.6~Ethics}
\dimrow{D6. NFRs for ML}
  {\cite{horkoff2019}}
\coderow{6.1~NFR trade-offs
  \textperiodcentered{} 6.2~NFR redefinition}
\addlinespace[2pt]
\bottomrule
\end{tabular*}
\end{table}

\subsubsection{First coding cycle}
\paragraph{Deductive coding}
The first author assigned each excerpt one \emph{primary} code (the dominant interpretation), optional \emph{secondary} codes, a confidence level, and an analytic memo. Of the 23 codes, 18 were applied as primary codes in 177 entries, three only as secondary codes, and two (discrimination/bias and feedback/control) never.

\paragraph{Inductive coding}
The first author also coded inductively, without a starting codebook. After the fourth interview, we consolidated the codes, refined their definitions, anchored each in a representative quote, and applied later codes retroactively to earlier interviews. A final LLM-assisted check added six excerpts to existing codes (Section~\ref{sec:llm}). This yielded 20 inductive codes with 78 applications across 77 excerpts. By the number of interviews in which each appeared, nine codes were \emph{consolidated} (three or more), seven \emph{moderate} (two), and four \emph{provisional} (one). Two codes appeared only in the last two interviews (one in each), so we do not claim saturation.

\paragraph{Second-author review}
The second author reviewed the annotated transcripts of all six interviews, assessing the deductive and inductive code assignments and searching for missed applications and new codes. The review raised 41 items (10 additions, nine removals, and 22 modifications), discussed in two reconciliation meetings, after the third and the sixth interviews. By consensus, 19 items changed the final coding, sometimes differently from the original proposal.

\subsubsection{Second coding cycle}
Using \emph{pattern coding}~\cite{saldana2009coding}, the first author grouped inductive codes that jointly explained a phenomenon into nine overlapping groups, checking each relation against the excerpts themselves. We crossed each group with the primary deductive codes on the same excerpts, classifying each relation as \emph{complements}, \emph{adds a dimension}, \emph{contrasts}, or \emph{no sustained relation}; all 18 primary codes related to at least one group. Merging groups with the same interpretation produced six candidate patterns, each a claim traceable to its excerpts. In a joint session (Google Meet and a Miro board), the second author reviewed the nine groups and then the six patterns, discussing their interpretations, overlaps, and boundaries, and the authors consolidated them by consensus. While writing the manuscript, the first author identified a seventh pattern, on requirements emerging in response to model outputs, from previously coded excerpts. Each of the seven patterns (A--G) documents its supporting participants and divergent evidence.


\subsection{Episode Consolidation}
\label{sec:consolidacao}

We consolidated the 17 EPs with a three-step protocol fixed on October 27, 2026. Each step used a separate LLM session, and the first author reviewed all outputs. First, the LLM located 94 discussion segments and extracted 106 candidate utterances with their preceding questions and suggested effects (\emph{confirms}, \emph{complements}, \emph{corrects}, \emph{contests}, or \emph{does not remember}). Only participants' statements counted as evidence; agreeing with a leading question did not. Twenty-one utterances informed only the gap analysis, and the first author revised 13 effects and one coverage classification. Second, we compared this evidence with five documentary analyses and the normalized items. The LLM proposed mergers, reformulations, and 59 candidate gaps, of which the first author kept 49 as complements or associations. Mergers required evidence of a shared transformation, not just thematic similarity.

Third, the LLM drafted consolidated episodes and proposed confirmation levels: \emph{high} (direct, specific support without unresolved contradiction), \emph{medium} (partial or qualified support), \emph{low} (weak, indirect, or contradictory support), and \emph{not assessable} (insufficient interview coverage). Documentary evidence could not raise these levels. The first author revised 10 of 13 drafts and raised C5, C8, and C10 from medium to high after restricting their claims to direct, uncontradicted evidence, although C10 relies mainly on one participant. The result is 12 sequential episodes (C1--C12) in five phases plus one cross-cutting episode (T1). Three mergers combined EP03+EP05, EP06+EP07, and EP12+EP14+EP15, and the first two phases became Phase~I; all 17 EPs remain traceable.

\subsection{Use of Large Language Models}
\label{sec:llm}
We used Claude Opus~5.5 (Anthropic), with maximum reasoning effort, for coding review (October 28, 2026) and episode consolidation (October 29, 2026). After inductive coding, it suggested six additional excerpts for existing codes, all accepted after review. During consolidation, it located segments, extracted utterances, suggested effects, cross-referenced documents, identified gaps, proposed mergers, and drafted episodes and confirmation levels. It also helped produce the trajectory figure (Fig.~\ref{fig:trajectory}) and the consolidation memo from the final spreadsheet. The first author reviewed all outputs and made all final decisions. A script checked quotations against the transcripts. Names were removed before the coding review, and all reported quotations are anonymized. 

\subsection{Ethical Considerations and Data Availability}
\label{sec:ethics}
All participants consented to recording and to the use of anonymized quotations. Names were removed before the LLM-assisted review, although context may still permit identification. Project artifacts and full transcripts cannot be released for confidentiality reasons. The replication package includes the interview guide, codebooks, normalized items, preliminary and consolidated episodes, the dated protocol, LLM prompts, and shareable analysis records.


\section{Results}
\label{sec:results}

The pipeline's requirements existed but were never written down: ``We never formalized this at any point'' (E02). Artifacts recorded features, criteria, and constraints, but no integrated specification, so we reconstructed the requirements from the documented demands, decisions, and results (Section~\ref{sec:req-def}).

\begin{figure*}
  \centering
  \includegraphics[width=0.9\textwidth]{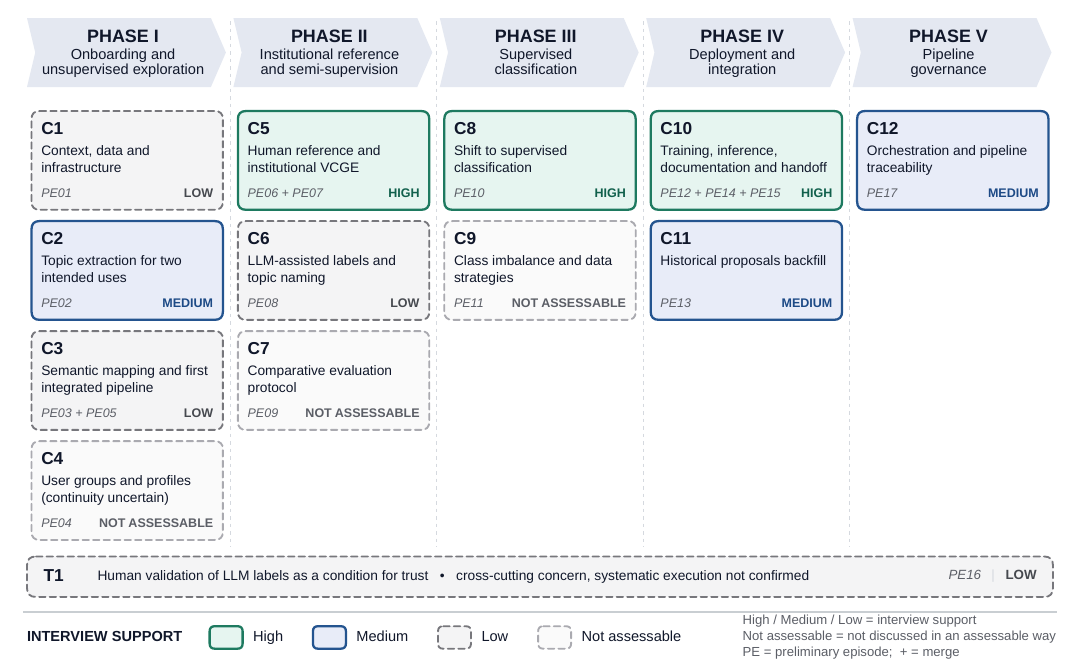}
  \caption{Consolidated trajectory of requirements evolution in the ML project, reconstructed from documentary artifacts and interviews.}
  \label{fig:trajectory}
\end{figure*}

Fig.~\ref{fig:trajectory} shows the consolidated trajectory: 12 sequential episodes (C1--C12) in five phases, from unsupervised exploration through institutional references and supervised classification to integration and governance, plus a cross-cutting episode (T1) on the recurring need to validate LLM-produced labels, whose systematic execution we could not confirm. Interview support was \emph{high} for C5, C8, and C10; \emph{medium} for C2, C11, and C12; \emph{low} for C1, C3, C6, and T1; and \emph{not assessable} for C4, C7, and C9. These levels measure neither documentary evidence nor implementation success.

We report the seven patterns (A--G) by RQ (Table~\ref{tab:padroes}); they describe related aspects of one process, not a causal sequence. Patterns A and B rest mainly on inductive codes, whereas G matches the deductive codes from the RE for ML literature most closely. We flag evidence from other AI fronts or from a single interview, and mark with~\textsuperscript{\dag} excerpts in which the participant addresses the first author (Section~\ref{sec:posicao}). The replication package traces patterns to codes and evidence.

Table~\ref{tab:profile} profiles the 95 requirements and criteria reconstructed from the artifacts. Only 38 (40\%) were formalized in an issue or meeting note; the rest were tacit. Two thirds first appear in Phases~I and~IV, when issues framed the work; Phases~II and~III left few records, and the requirements tied to the institutional reference were never formalized (Pattern~C). Six of the seven environmental constraints come from one Phase~IV issue on the production database; hardware and contracts appear only in passing (Pattern~F). No success criterion was formalized (Pattern G), and the interviews add 14 requirements, four of them environmental, and 7 criteria beyond the 95.

\begin{table}
\centering
\caption{Requirements and criteria reconstructed from the artifacts.}
\label{tab:profile}
\begin{tabular*}{\columnwidth}{@{\extracolsep{\fill}}lrrrrrrrrr}
\hline
 & & & & & \multicolumn{5}{c}{Phase of first appearance} \\
\cline{6-10}
 & $n$ & Form. & Tacit & Env. & I & II & III & IV & V \\
\hline
FR    & 21 & 13 &  8 & 1 &  7 &  4 & 2 &  7 & 1 \\
NFR   & 12 &  6 &  6 & 2 &  5 &  3 & 1 &  2 & 0 \\
AC    & 23 & 13 & 10 & 2 &  9 &  3 & 2 &  8 & 1 \\
EC    & 22 &  6 & 16 & 1 &  9 &  4 & 3 &  6 & 0 \\
SC    & 17 &  0 & 17 & 1 &  6 &  2 & 1 &  6 & 1 \\
\hline
Total & 95 & 38 & 57 & 7 & 36 & 16 & 9 & 29 & 3 \\
\hline
\end{tabular*}
\par\smallskip
\parbox{\columnwidth}{\footnotesize Form.: stated explicitly in an issue or meeting note; Tacit: all others. Env.: environmental constraints, marked across formalized and tacit items (Section~\ref{sec:req-def}). One NFR (RNF12) and one SC (CS05) have no consolidated episode.}
\end{table}


\begin{table}
\centering
\caption{Patterns by research question and the consolidated episodes used for triangulation.}
\label{tab:padroes}
\footnotesize
\setlength{\tabcolsep}{4pt}
\begin{tabular}{@{}c>{\raggedright\arraybackslash}p{0.63\columnwidth}l@{}}
\toprule
 & \textbf{Pattern} & \textbf{Episodes} \\
\midrule
\multicolumn{3}{@{}l}{\textit{RQ1: from demands to requirements}} \\
A & AI entered as the team's offer, before its use was defined & C1--C3 \\
B & Endorsement did not become sponsorship & C10--C12 \\
C & The classification reference had no convergent owner and reoriented the task & C3, C5, C8 \\
\addlinespace
\multicolumn{3}{@{}l}{\textit{RQ2: requirements and technical decisions}} \\
D & Values and resources acted as tacit requirements, met unevenly & C1, C5, C8, C10 \\
E & Classification requirements surfaced in reaction to concrete outputs & C2, C3, C5, C7--C9 \\
F & Operational requirements surfaced in development and integration & C10--C12 \\
\addlinespace
\multicolumn{3}{@{}l}{\textit{RQ3: acceptability}} \\
G & Acceptance rested on judgment, and evaluation in use was missing & C5, C7, C10, T1 \\
\bottomrule
\end{tabular}
\end{table}

\subsection{RQ1: How Institutional Demands Became Requirements}
\label{sec:rq1}

\subsubsection{Pattern A: AI entered as the team's offer, before its use was defined}
\label{sec:pa}

The need to analyze citizen contributions at scale was recognized but did not translate into an institutional demand for AI. According to the requesting body's director, AI ``does not enter at the level of political definition'' and was absent from the first work plan; it was a team initiative that the body welcomed (E03). The product team called AI a ``self-contained'' demand, unlike features frequently validated with the requesting body (E04). Work advanced from opportunities the team identified: facing the broad goal of increasing engagement, E02 proposed to ``break it into parts'' and explore solutions, and the coordination linked proposal classification to other Decidim deployments (E05). Moderation is an exception: the need was acknowledged, although the team translated it into an ML solution (E04, E05).

Endorsement did not settle who would consume each output. E03 reported an external expectation that the platform have AI, ``almost a certificate that you are riding the wave,'' and said the technical team chose the tools and the body ``just gave the OK.'' Intended uses included informing managers and recommending proposals to citizens, but the feedback to citizens was never designed, the topics came to be seen as ``maybe'' useful to managers (E02), and the classification was evaluated before anyone defined how a service would display it (E01). The trajectory shows the same precedence: the onboarding (C1) mapped data, tools, and environment constraints while the model's goal and integration remained underspecified; topic extraction was assessed with technical metrics before any service or acceptance threshold existed (C2); and extraction and categorization were integrated into a pipeline with no evidence that the categories helped managers (C3). C1 and C3 have low interview support, so this part rests mainly on technical records. For the coordination, classification and RAG-based reporting became clearer products only at a workshop ``almost two years later'' (E05).

The artifacts point the same way. Tracing each FR and NFR to the earliest artifact that states it, all were first recorded by delivery-side technical roles: ML leadership dominated Phase~I (8 of 12 entries), the AI team led Phases~II and~III, and data and infrastructure roles joined only in Phases~IV and~V, with deployment. None was first recorded by the requesting body, the product team, or moderation, although this partly reflects a corpus without government-side artifacts (Section~\ref{sec:threats}). Authorship shows who wrote a requirement down, not who conceived it, but it matches the interviews: the team formulated requirements for a use others had only endorsed. The finding is not an absence of purpose but an inversion between component and service: exploring an ML capability preceded a precise definition of its use and recipients.


\subsubsection{Pattern B: Endorsement did not become sponsorship}
\label{sec:pb}

We distinguish \emph{endorsement}, approving an initiative, from \emph{sponsorship}, the continued action of someone who can prioritize deliveries and remove institutional obstacles. AI was endorsed but stayed peripheral. ``Machine learning was not a political priority'' next to keeping the platform running and migrating it (E04); the product ``doesn't need AI to work'' (E02); and even as an official front, AI ``never came in as a delivery priority'' (E05). For the former director, the main concerns were other features and security, while AI was still ``very much running in the lab'' (E03).

Priority also varied with politics. E04 associated stronger delivery pressure with an administration that emphasized innovation, but E03, part of that administration, said AI was not central to it; the accounts do not support stable sponsorship even then. Later, the platform's turn to participatory budgeting reduced AI's space (E02), and the electoral calendar blocked the deployment of a solution the team considered ready (E01), although this readiness was never measured.

The infrastructure dependency shows the consequences. The coordination reported a long negotiation to obtain GPUs from the state-owned company that hosts the platform, contrasting it with the WhatsApp integration, which reportedly secured a contract quickly once it became a new administration's ``top priority'' (E05). For E05, AI lacked the political interest needed to unlock it; E02 likewise called for someone to bridge AI and the product. This link comes mainly from an interested party and does not show that the obstacles were only political. Still, the accounts converge on a mechanism: ML delivery depended on decisions, resources, and teams over which the technical team had little authority, and approval did not ensure that these dependencies would be resolved. Patterns~F and~G show how they constrained operation and evaluation in use.


\subsubsection{Pattern C: The classification reference had no convergent owner and reoriented the task}
\label{sec:pc}

Defining the categories was a constitutive requirement: without deciding what each class represented, there was no shared reference for labeling proposals, training the model, or interpreting its outputs. According to E05, the category dictionary ``went through several hands'' (the requesting body, the team, and moderation) and, when it seemed close to done, was rejected: ``that's not at all what I want, from a business point of view.'' The reference requested from the domain side also arrived slowly (E05). The problem was not only insufficient labels but disagreement about what the categories were for and who had the authority to define them. In C3, the semantic-mapping and integration experiments show that the technical coverage of the associations did not establish whether the categories fit the domain.

The VCGE offered an institutional way out. E03 stated that the requesting body suggested it when the team asked for a vocabulary, without considering it good, only ``the best parameterized one available at that moment.'' E01 also claimed to have suggested it, since a reference valid ``for any agency, on any topic'' was missing. The interviews do not establish who suggested it first, but they reveal distinct justifications: availability (E03), generality across the platform (E01, E05), legitimacy to explain where the themes came from (E01), and managers' familiarity with the nomenclature (E04). The team did not create the VCGE but appropriated and adapted it. C5 documents the comparison with human classifications and the later VCGE guidance as distinct functions, joined by the pursuit of alignment with public policy. E04 gathered classifications from several people, harmonized the codes, and added public-policy elements the VCGE lacked. In a different front, hate-speech detection, E02 described relabeling ``100\%'' of a dataset because annotators disagreed; this is not the thematic dataset, but it shows a similar mechanism of technically constructing the reference.

The institutional reference then helped reorient the method. E04 linked the move to supervised learning to tying results to a nomenclature managers knew; E02 stressed the difficulty of objectively evaluating unsupervised clusters and the possibility of computing metrics against reference labels. These justifications are complementary, not a single trigger, and C8 documents the transition to supervised classifiers and predictive metrics without an exclusive cause or evidence of multi-label output. E01 added that the VCGE ``doesn't change the way of evaluating'' on the domain side; it mainly helped to ``unlock the project.'' The taxonomy thus provisionally settled legitimacy and output specification without showing that its categories suited the language of those who wrote the proposals. The difficulty of establishing shared classes echoes specification problems in RE for ML~\cite{habibullah2023}; here, the reference also had to confer institutional legitimacy.



\subsection{RQ2: How Requirements and Technical Decisions Shaped Each Other}
\label{sec:rq2}

The relationship ran in both directions. Values and resources constrained decisions from the outset (D), while classification criteria surfaced from model outputs (E) and operational requirements from development and integration (F).

\subsubsection{Pattern D: Values and resources acted as tacit requirements, met unevenly}
\label{sec:pd}

The political guideline on technology, stated by E03, was to avoid coupling a ``ready-made market solution'' to the platform. Grounded in open knowledge, data control, and technological autonomy, it favored an in-house, ``small-scale'' model trained on project data (E03). This choice bounded the solution and required infrastructure able to run an open stack, tying the pipeline to the dependency in Pattern~B. C1 records these environment limitations early, though never as formalized NFRs; Pattern~F shows how they became a CPU-inference requirement. The principle was applied unevenly: in another front, a prototype built on commercial APIs set the benchmark for the in-house models (Pattern~E).

Other conditions came from the purposes of participation. Outputs had to be intelligible to managers and society: E01 called for ``less bureaucratic language'' and ``more citizen language,'' and E03 criticized the VCGE as ``the language of the administration, not the language of the citizen.'' C5 records the adoption of the institutional reference with the caveat that its fit to citizens' language remained contested. The classification also had to cut across processes and agencies, since Decidim offered no global categorization (E05), and to represent proposals spanning several policies: ``almost every proposal involves several policies'' (E03). Yet the team limited the output to two categories for simplicity, although a proposal could belong to ``three, four categories'' (E04), and C8 shows no multi-label output. These conditions competed: autonomy and generality drove effective choices, citizens' language stayed in tension with the taxonomy, and transversality was only partly met. No interview reports validation with citizens.


\subsubsection{Pattern E: Classification requirements surfaced in reaction to concrete outputs}
\label{sec:pe}

Participants could not always anticipate what would make the classification useful; some expectations became explicit only when they examined concrete results. The rejected category dictionary (Pattern~C) exposed disagreement about what made a category useful. In the unsupervised approach, the team realized that the generated themes could impose ``a very large cognitive load'' on managers (E04), which strengthened the need for understandable, stable categories and contributed to the move to supervised learning. In C2 and C3, experiments already exposed trade-offs among the coherence, diversity, and number of topics, and between coverage and over-association, with no institutional criterion to resolve them. When domain experts examined a sample, they judged the classification ``still very far off'' (E04), a verdict that identified inadequacies but set no threshold. C7 and C9, not discussed in the interviews, show the same: C7 compared internal and external metrics with no institutional threshold (Pattern~G), and in C9 class reorganizations improved metrics without confirmed usefulness to the domain, while synthetic data augmentation brought no relevant gain.

Reactions could also define positive references. In the Climate Plan systematization, reported only by E01, the ministry's agreement with the themes of a prototype built on commercial APIs turned that output into a target for the in-house models: ``our models have to at least handle this.'' This suggests specification by example, in tension with the autonomy principle (Pattern~D), though it does not confirm the mechanism for the thematic classification.

One interview showed both the role of examples and a methodological limit. Facing a table shown by the interviewer, E03 questioned the single label (``But do they only accept one tag?''), examined how ``domestic violence'' related to ``gender violence,'' and lacked ``clarity on how it would be used.'' Since E03 said multi-label output had been discussed before, the reaction shows a concrete output letting an existing expectation be discussed and tested, possibly influenced by the interviewer.

Overall, outputs judged inadequate prompted revisions, and outputs judged ``plausible'' could become benchmarks. The episodes document successive adjustments of criteria and approaches, without formal validation cycles. This pattern concerns classification criteria; Pattern~C covers their reference and Pattern~F operational requirements.


\subsubsection{Pattern F: Operational requirements surfaced in development and integration}
\label{sec:pf}

Operational requirements surfaced when the component met data, teams, and environments. According to E06, requests reached development without enough technical analysis, and refinement ``happened at the development level'' whenever a problem required ``going back to the definition.'' In data preparation, business rules ``unknown to us and even to the product team'' had been applied inconsistently and required extra cleaning (E06). The model was trained on CSV files delivered before automated ingestion matured, and E06 stated that, ``the way it was done,'' the training data was not traceable; later pipelines may differ.

In the integration with engineering, discussions focused on inputs, outputs, and formats: ``The inner logic of how you did it wasn't that necessary'' (E06\textsuperscript{\dag}). In one mismatch, the product expected three categories with their scores but received one category with three scores (E06). In the RAG front, an architecture designed ``at the algorithmic level'' had to be reworked to run as a service (E05). Dependencies among teams also shaped what became deliverable: to avoid another bottleneck, the ML team began to expose the model as an API and to handle part of its integration (E04). C10 documents hand-off, documentation, training, publishing, and inference as requirements of a service for other teams.

The environment added further requirements. Inference had been developed on lab GPUs, but the external production infrastructure had none, so deployment required CPU inference (E06), which the coordination also saw as a way to avoid amending the contract (E05); this is a hardware and governance constraint, with no evidence of degraded accuracy. Classifying proposals at creation time also meant classifying existing ones, which required writing results to an auditable production table for which the team lacked credentials (E06); C11 makes this need explicit, with no evidence that the update took place. Since the feedback service to citizens was never designed, it remained unclear how participants could be contacted under data-protection law and whether one or three suggested categories would serve better (E02). Later, C12 documents the adoption of Airflow and MLflow for orchestration, versioning, and traceability, which still did not reach the production database governed by another organization.

Part of the requirements thus became visible only when the solution met real data, interfaces, and resources managed by other actors. Some changes during development, however, corrected defects, inconsistencies, or insufficiently refined decisions and were not emergent requirements.



\subsection{RQ3: How Acceptability Was Established}
\label{sec:rq3}

\subsubsection{Pattern G: Acceptance rested on judgment, and evaluation in use was missing}
\label{sec:pg}

We found no shared quantitative threshold for accepting the classification. E01 expected ``a high hit rate'' but admitted that ``we hadn't defined what that would be,'' and E04 observed that presenting ``the F1'' did not tell stakeholders what the result meant. Domain-side assessment examined examples, asking ``does this proposal really belong to this theme? Does it make sense?'' (E01); it could involve samples and several people, but never became a measurable readiness criterion (E04). For E02, the difficulty of objectively evaluating clusters was one more reason to adopt supervised learning, where metrics against reference labels became possible. C7, not discussed in the interviews, shows that a comparative protocol of internal and external metrics did not by itself yield an institutional acceptance criterion.

\begin{figure}[t]
  \centering
  \includegraphics[width=\columnwidth]{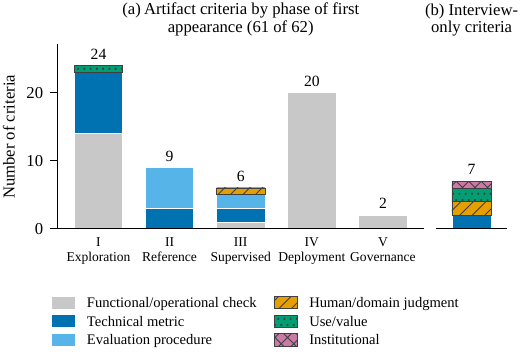}
  \caption{Nature of the acceptance, evaluation, and success criteria:
  (a)~artifact criteria by phase of first appearance; (b)~criteria found
  only in interviews. SC05 has no consolidated episode and is omitted.}
  \label{fig:criteria-profile}
\end{figure}

The artifacts show the same imbalance (Fig.~\ref{fig:criteria-profile}). Most of their 61 criteria are functional or operational checks and technical metrics; only one records a domain judgment, none an institutional condition, and all criteria first appearing in Phases~IV and~V are operational checks. Judgment-based criteria surface mainly in the interviews, and none of the 17 success criteria was formalized. What decided acceptance thus went largely unrecorded.

Judgment took different forms. The technical team checked coherence and compared results with human classifications; managers assessed whether the categories made sense for the domain; and the political assessment weighed whether the solution was ``mature enough'' to enter production without compromising the platform's credibility (E05). These expectations could diverge: E03 wanted to review proposals individually, while E04 described a classification that still needed revision after expert testing. Verification also covered input data, as people familiar with the business rules spotted inconsistencies (E06). T1, with low interview support, records human validation of automatic labels as a recurring concern but shows no systematic review of LLM-produced labels.

Evaluation in final use remained limited. In moderation, the team expected to compare predictions with moderators' decisions, but ``we never got feedback from them'' (E02); for the classification, E01 stated that the feature's usefulness was never evaluated. ``Production'' also meant different things: running in the lab differed from being incorporated into the platform maintained by the state-owned company, with access to its database (E02, E06). Neither C10 nor C12 shows that the supervised classifier was deployed in production. Our claim is therefore limited: neither the interviews nor the episodes show the classification's usefulness being validated by its end recipients, in part because of the unresolved dependencies of Pattern~B. The difficulty of turning metrics into criteria that customers understand resembles the problem described by Vogelsang and Borg~\cite{vogelsang2019}; here, political endorsement added an acceptance condition beyond predictive performance.


\subsubsection{What did not appear: bias and citizens}
\label{sec:ausencias}

Two absences bound these results. First, we found no substantive discussion of performance differences across groups of citizens or ways of writing proposals. The bias and discrimination code (1.3) was never applied, and ethics (5.6) appeared only as a secondary code. In E04, the mention of ``bias'' concerned possible bias in the human judgment of outputs, not discrimination by the system. Because the interview guide did not specifically address algorithmic bias or discrimination, this absence does not show that such risks were ignored in the project.

Second, citizens were not interviewed and, in the coded excerpts, do not help define or evaluate the categories. Their perspective appears only indirectly, through concerns in understandable language (E01, E03), multiple categories (E03), and feedback to society (E01, E03); E05 also mentioned using the classification to guide managers and recommend activities to citizens. The accounts do not show that the categories or the results were validated by the people who submitted the proposals. This absence is limited to the evidence analyzed and cannot be extended to other project activities.

\section{Discussion}
\label{sec:discussion}

\textbf{Requirements were not missing; they were distributed.} RE for ML has documented that data scientists rarely receive explicit requirements~\cite{vogelsang2019}. Our case suggests a different reading of that absence: the pipeline's requirements existed, but they were spread across issues, notebooks, presentations, and people, and only 40\% were ever formalized. They took shape less through elicitation than through approval, legitimacy, values, choices of reference, and the capacity to prioritize. Experiments and model outputs acted as elicitation instruments: stakeholders who could not state what a useful category was could reject one when they saw it (Pattern~E), and a plausible prototype became a target. This suggests treating experimentation as part of the requirements process, as agile RE does with prototyping and as requirements and architecture co-evolve~\cite{cao2008,nuseibeh2001}. In practice, this means recording requirements where they surface, for instance as an explicit ``requirement revealed'' field in experiment reports, so that the 57 tacit items we had to reconstruct are captured as they appear. It also suggests that artifact mining, as we did here, can recover requirements that no one wrote down.

\textbf{Public-sector ML adds requirements that RE for ML catalogs do not name.} The central requirement, the classification categories, had no owner, and the taxonomy that filled the gap was chosen for its availability and institutional legitimacy, and it never reflected citizens' language (Pattern~C). Legitimacy thus behaved as a quality requirement of its own, absent from current NFR agendas for ML~\cite{horkoff2019}, and it competed with intelligibility for citizens, the people who write the input but do not consume the output. Endorsement without sponsorship (Pattern~B) left this and other trade-offs unresolved and every dependency on third-party infrastructure unowned. Human-centered RE for AI calls for eliciting user needs~\cite{ahmad2023}; in this setting, a prior question is \emph{who} the users are and who has the authority to decide for them. Public-sector ML projects could make these roles explicit before modeling starts: an owner for the reference vocabulary, a sponsor for deployment dependencies, and a channel for those who produce the input.

\textbf{Acceptance needs to be engineered.} As in industrial settings~\cite{vogelsang2019,nahar2022,paleyes2022}, metrics did not translate into stakeholder judgment, and moving the model into the product exposed requirements that modeling had not anticipated. Our case adds two layers: a political judgment of the risk to the platform's credibility, and an ambiguous notion of ``production'' tied to infrastructure and contracts outside the team's control (Patterns~F and~G). With no shared threshold and no evaluation in use, the pipeline could be \textit{technically ready and institutionally unaccepted} at the same time. A lightweight remedy is to agree early on a sample-based review protocol with domain experts, possibly expressed as behavioral tests~\cite{ribeiro2020}, with an explicit acceptance threshold and a plan for evaluation in use. For researchers, the gap between single-point accounts and what we observed over 15 months argues for more longitudinal studies of ML requirements, and for including the managers and citizens our study could not reach.

\subsection{Limitations}
\label{sec:threats}
Our study has several limitations. First, the requirements, criteria, and episodes were not recorded as such; we reconstructed them, and others might have segmented them differently, although each item links to its original artifacts. The deductive codebook may also have drawn our attention toward what its four sources describe, which inductive coding offsets. Second, showing episodes to participants may have steered their answers, so an utterance confirmed an episode only if it added the participant's own elements (Section~\ref{sec:consolidacao}); only one reaction in Pattern~E occurred in an interview. Third, the study is retrospective and prone to recall bias, and seven of the 13 consolidated episodes rest mainly on artifacts, in which the government side is underrepresented; participants may also have softened criticism of the government or colleagues before an interviewer from the team. Fourth, the first author was a team member who conducted the interviews and most of the analysis; the second author reviewed all coding (19 of 41 items led to changes; we report no agreement coefficient, as the review aimed to refine the interpretation) and the six candidate patterns, but not the seventh, and we did no member checking. LLM outputs may not be reproducible, although the first author reviewed every suggestion. Fifth, we did not interview citizens or ministry managers, and the guide did not ask about algorithmic bias, so both absences (Section~\ref{sec:ausencias}) may partly reflect our design. Finally, this is a single case: we aim at analytical generalization~\cite{yin2018} to public-sector ML projects with similar conditions, namely no product owner, input producers distinct from output consumers, and third-party infrastructure. Quotes were translated from Portuguese, and artifacts and transcripts cannot be released, which limits independent verification.

\section{Related Work}
\label{sec:related}

\textbf{RE for ML.} Data scientists rarely receive explicit requirements, and customers struggle to state performance targets~\cite{vogelsang2019}; practitioners of ML-based perception systems cannot specify requirements upfront and rely on operational design domains instead~\cite{habibullah2023,habibullah2024}. A survey of 188 practitioners found that RE for ML is done mostly by project leaders and data scientists, that notebooks are the main documentation format, and that unclear goals and low customer engagement are common problems~\cite{alves2023}. Research agendas cover non-functional~\cite{horkoff2019} and human-centered requirements~\cite{ahmad2023}, and mapping studies rank specification, explainability, and the gap between ML engineers and end users as the most prevalent challenges~\cite{villamizar2021,habiba2024}.  These studies capture RE at one point in time, through interviews and surveys, or treat it as a specification step before modeling. Our findings agree with theirs, notebooks included~\cite{alves2023}, but we follow one pipeline for 15 months and reconstruct from dated artifacts \emph{when} and \emph{from what} each requirement surfaced: most emerged from experiments and integration after modeling had started, which neither single-point accounts nor upfront specification can observe.

\textbf{SE challenges of ML systems.} Industrial studies and reviews catalog challenges across the ML workflow, from data management to deployment and maintenance~\cite{amershi2019,lwakatare2019,lwakatare2020,paleyes2022}, including hidden technical debt such as glue code and undeclared consumers~\cite{sculley2015}, friction between data scientists and software engineers when models enter products~\cite{nahar2022}, and data cascades caused by undervaluing data work in high-stakes domains~\cite{sambasivan2021}. Our operational requirements (Pattern~F), from mismatched output formats to untraceable training data, resemble these challenges. These works, however, treat them as engineering problems that follow requirements; we show them as a \emph{source} of requirements and add institutional causes, such as contracts, third-party infrastructure, and missing sponsorship, that industrial settings rarely face.

\textbf{AI in the public sector.} Public administration research has mapped AI applications and challenges for governments~\cite{wirtz2019,desousa2019}, called for a prior debate with society on the use of AI~\cite{desousa2019}, and shown that stakeholders frame the challenges of adopting AI differently~\cite{sun2019}. These studies take a policy view and treat the AI system as given. We look inside the engineering of one system and show how such divergent framings become concrete requirements: a taxonomy chosen for legitimacy, a political judgment of credibility as an acceptance criterion, and citizens absent from defining the categories.


\section{Conclusion}
\label{sec:conclusion}
RE for ML often assumes that requirements are elicited upfront and then drive design and evaluation. Following a public-sector ML pipeline for 15 months, we found that its requirements existed but nobody wrote them down. AI entered as the team's offer, endorsed but never sponsored; the classification categories had no owner and were settled by a government taxonomy; and most requirements surfaced from model outputs and from integration with data, teams, and third-party infrastructure. Acceptance rested on layers of judgment, with no shared threshold or evaluation in use. Experiments, model outputs, and institutional artifacts were how these requirements came to exist.

\bibliographystyle{IEEEtran}
\bibliography{references}

\end{document}